\documentclass{PoS}
\newcommand{\ssr}{Space Science Reviews}

\newcommand{\apjl}{ApJ}
\newcommand{\apj}{ApJ}
\newcommand{\prl}{Phys. Rev. Lett.}

\title{Inductive spikes and gamma-ray flares from the Crab Nebula}

\ShortTitle{Gamma-ray flares}

\author{\speaker{John G. Kirk}\\
        Max-Planck Instute for Nuclear Physics, Saupfercheckweg 1, 
69117 Heidelberg\\
        E-mail: \email{john.kirk@mpi-hd.mpg.de}}

\author{Gwenael Giacinti\\
Max-Planck Instute for Nuclear Physics, Saupfercheckweg 1, 
69117 Heidelberg\\
        E-mail: \email{gwenael.giacinti@mpi-hd.mpg.de}}

      \abstract{The $\sim400\,$MeV flaring emission from the Crab
        Nebula is naturally explained as the result of an abrupt
        reduction in the mass-loading of the pulsar wind. Very few
        particles are then available to carry the current required to
        maintain wave activity, causing them to achieve high
        Lorentz factors. When they penetrate the Nebula, a tightly
        beamed, high luminosity burst of hard gamma-rays results,
        with characteristics similar to the observed flares. This
        mechanism may operate in other powerful pulsars, such as
        J0537$-$6910 (PWN\,N\,157B), B0540$-$69, B1957$+$20 
and J0205$+$6449 (3\,C\,58).}

\FullConference{7th Fermi Symposium 2017\\
		15-20 October 2017\\
		Garmisch-Partenkirchen, Germany}

\begin{document}

\section{Introduction}
One of the most interesting discoveries of the AGILE and Fermi
satellites is the rapid variability of the emission at hundreds of
MeV from the Crab Nebula \cite{Agileflares11,Fermiflares11,buehleretal12}. 
Numerous suggested explanations have
appeared in the literature, but, to date, a consensus is lacking. (For
a review, see \cite{buehlerblandford14}.)  In a recent {\em Letter} we
suggested that inductive, radial acceleration of particles
in the pulsar wind is responsible \cite{kirkgiacinti17}. Here we give
a brief description of this mechanism, and discuss its relevance in
other objects.

\section{Inductive acceleration}
Crucial to our mechanism is the idea that the pulsar wind is
predominantly an electromagnetic wave, that carries along a relatively
small number of charged particles, specifically, electron-positron
pairs. Close to the pulsar, the density of these particles is high
enough to allow the wave to be described by the equations of MHD.
Consequently, we assume the wave is launched as a mildly supersonic
flow that contains an oscillating, frozen-in magnetic field --- a
configuration commonly referred to as a {\em striped wind}. 
This is a widely
adopted assumption (see \cite{porthetal17} for a more detailed
discussion), 
which has some observational support \cite{mclaughlinetal04}. 
An inescapable consequence is that, at some radius, the density of 
charges becomes too small to justify the MHD approximation. 
At this point, the flow begins to accelerate 
\cite{lyubarskykirk01,kirkskjaeraasen03,spruitdrenkhahn04,kirkmochol11,zrake16}. 
In terms of the basic wind parameters:
\begin{eqnarray}
a_{\rm L}&=&\left(e^2 L_{\rm s.d.}/m^2c^5\right)^{1/2}
\,=\,3.4\times10^{10}L_{38}^{1/2},
\end{eqnarray}
where $L_{\rm s.d.}=L_{38}\times 10^{38}\,\textrm{erg\,s}^{-1}$
is the spin-down power of the neutron star, and
\begin{eqnarray}
\mu&=&L_{\rm s.d.}/\left(\dot{N}_\pm mc^2\right),
\label{mudef}
\end{eqnarray}
where $\dot{N}_\pm$ is the rate at which electrons and positrons are
transported into the nebula by the wind, the magnetization parameter
is given by 
\begin{eqnarray}
\sigma(r)&=&\mu/\gamma(r) -1
\end{eqnarray}
where $\gamma(r)$ is the particle Lorentz factor,
which, like $\sigma$, is a function of radius $r$.  The \lq\lq
multiplicity\rq\rq\ parameter $\kappa$ commonly used in modeling pair
production near the pulsar \cite{ceruttibeloborodov17} is roughly
$\kappa=a_{\rm L}/\left(4\mu\right)$ but this should not be
interpreted too literally, since a self-consistent model linking pair
production with the wind parameters is not yet available.

\begin{figure}
\includegraphics[width=15cm]{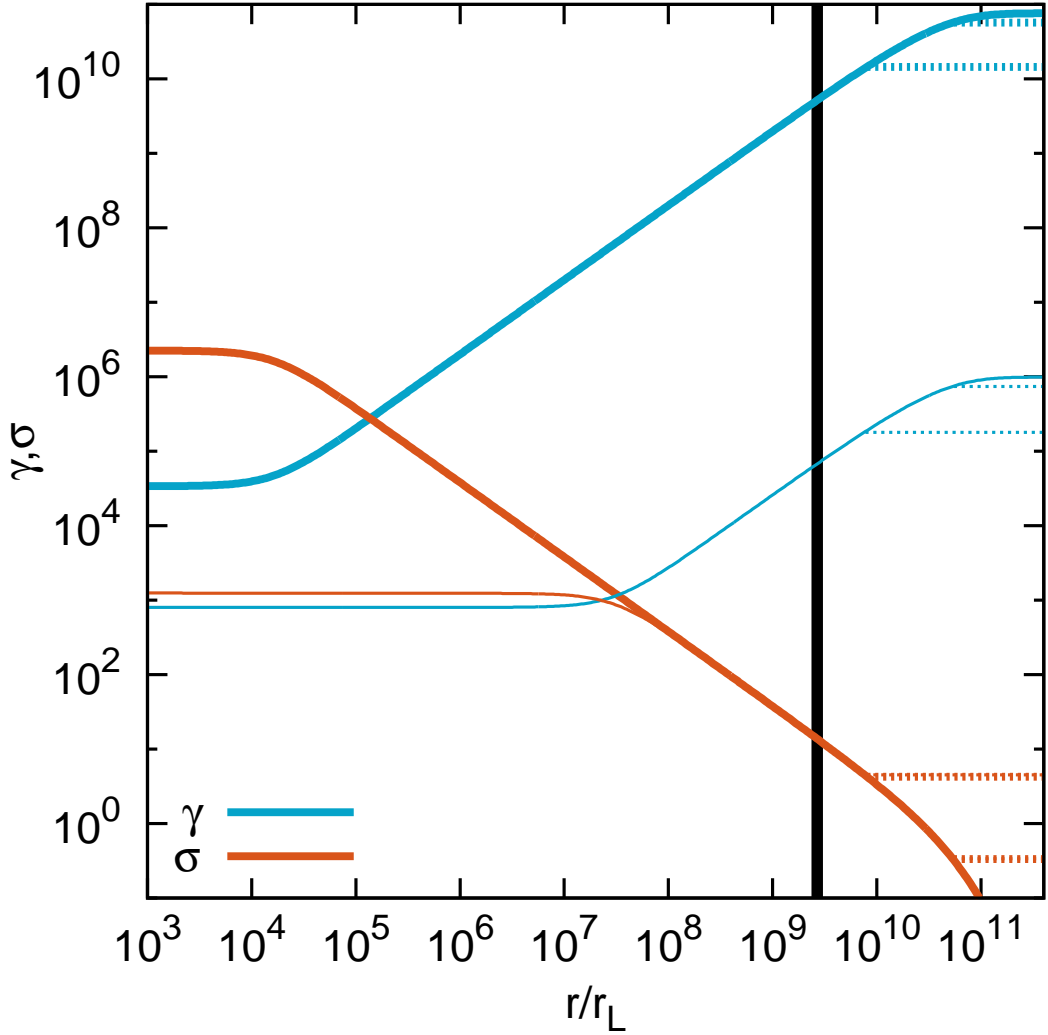}
\caption{\label{fig1}%
The magnetization 
parameter 
$\sigma$ and Lorentz factor $\gamma$  
for high density ($\mu=10^6$, thin lines) and low 
density ($\mu=a_{\rm L}$, thick lines), with 
$a_{\rm L}=7.6\times 10^{10}$ (corresponding to the Crab). 
The thick vertical line indicates the position of the 
termination shock in the Crab \cite{hesteretal02}.
}
\end{figure}

Based on a model with cold electron and positron fluids, an 
approximate solution in the acceleration phase 
can be found \cite{kirkmochol11}:
\begin{eqnarray}
u_\bot&\approx&1\qquad
\gamma\,\approx\,2\mu r/\left(a_{\rm L}r_{\rm L}\right)
\qquad \sigma\,\approx\,r_{\rm L}a_{\rm L}/\left(2r\right),
\label{accelerationzone}
\end{eqnarray}
where $u_\bot$ is 
the transverse component of the four velocity of the fluids (in units of $c$) and 
$r_{\rm L}$ is the light-cylinder radius. A full, numerical integration of the 
radial evolution equations (see \cite{kirkgiacinti17}) 
is shown in Fig.~\ref{fig1} --- the approximate solution
(\ref{accelerationzone}) is valid in the range 
$a_{\rm L}\gamma_{\rm L}/\mu\ll r/r_{\rm L}\ll a_{\rm L}$, where
$\gamma_{\rm L}$ is the initial Lorentz factor of the wind. 
Except exactly in the equatorial plane, the wind carries with it a
non-zero DC component, that cannot be converted into kinetic
energy. Thus, at finite latitude, 
the accelerating phase of the solutions in Fig.~\ref{fig1} 
terminates when the wave energy is exhausted, even though $\sigma>0$, 
and the wave subsequently 
proceeds at constant $\gamma$ and $\sigma$, as shown by horizontal, 
dotted lines for initial wave amplitudes equal to 
$10\,\%$ and $50\,\%$ of the DC component at launch. 

For the low-density wind, which corresponds to $\kappa\sim 1$, the electrons and 
positrons in the wind of the Crab pulsar achieve Lorentz factors of roughly 
$5\times10^9$ when they reach the termination shock, although, 
at this point, only 
about $10\,\%$ of the wave energy has been converted into particle kinetic 
energy.

\begin{figure}
\includegraphics[width=15 cm]{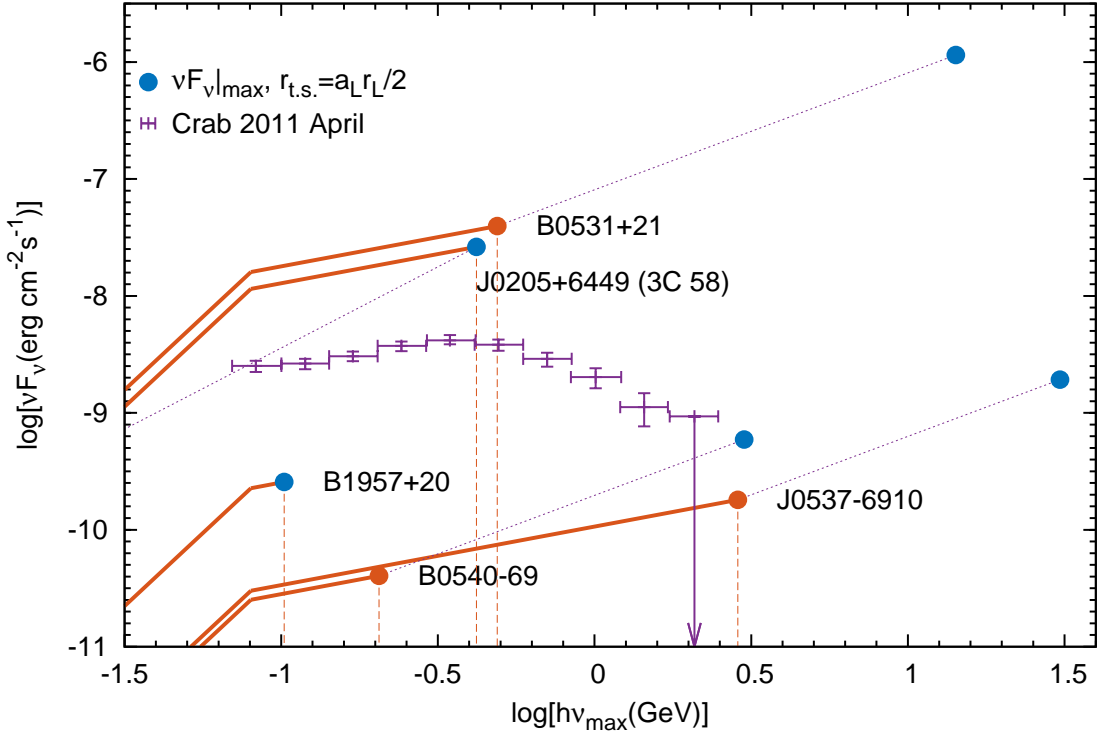}
\caption{\label{fig2}%
The predicted flare spectrum (solid lines), for the 
three most powerful known pulsars: the Crab 
(B0531$+$21), and two objects in the Large Magellanic Cloud, assuming 
a turnover at $h\nu_{\rm t}=80\,\textrm{MeV}$  
and a filling factor $f=1$.  
For B0531$+$21, J0537$-$6910 and B0540$-$69, dotted lines trace the locus of  
the peak flux as the position of the termination shock is varied between 
the observed value (orange dots)  and $a_{\rm L}r_{\rm L}/2$ 
(blue dots). For J0205$+$6449 (3C~58) and B1957$+$20 only the (optimistic) blue dots and
the corresponding spectra are shown.
Fermi observations of the powerful flare from the Crab Nebula 
in April~2011 are also shown --- 
points taken from Fig~6, epoch~7 of Ref.~\cite{buehleretal12}.
}
\end{figure}

\section{Gamma-ray flares}
On impacting the Nebula, a pocket of low density wind, with $\kappa\sim1$, 
injects radially directed, 
PeV electrons into the turbulent downstream magnetic field, whose strength
is roughly three times that in the wind, i.e., 
$B\approx3\times\left(2\pi m c/e P\right)\left(a_{\rm L}r_{\rm L}/r\right)$, where
$P$ is the pulsar period and $r_{\rm t.s.}$ is the radius
of the termination shock ($33\,$ms and $4.3\times10^{17}\,\textrm{cm}$
in this case). The resulting synchrotron emission peaks at 
\begin{eqnarray}
h\nu_{\rm max}&\approx&18 a_{\rm L}^2 \left(h/P\right)
\left[r_{\rm t.s.}/\left(a_{\rm L}r_{\rm L}\right)\right]
\nonumber\\
&\approx&500\,\textrm{MeV}
\label{numaxeq}
\end{eqnarray}
An electron initially radiating at $h\nu_{\rm max}$ is 
deflected by 
\begin{eqnarray}
\delta\theta\left(\nu\right)&\approx&\left(80\,\textrm{MeV}/h\nu\right)
\left(1-\nu/\nu_{\rm max}\right)\,\textrm{radians}
\label{divergence}
\end{eqnarray}
whilst cooling down to emit at $h\nu$.
Therefore, photons emitted between $400$ and $500\,$MeV by the electrons
injected into the Crab Nebula remain radially collimated in a cone
of opening angle about $2.5^\circ$. Assuming a cone of pulsar wind of solid
angle $\Omega$ contains low-density pockets with volume
filling factor $f$, the flux seen by an observer at distance $D$, 
whose line of sight lies inside this cone is
\begin{eqnarray}
\nu F_\nu&\approx&
f L_{\rm s.d.}/\left(8\pi\sigma\left(r_{\rm t.s.}\right) D^2\right)
\left(\nu/\nu_{\rm max}\right)^{1/2}
\label{predictedspectrum}
\end{eqnarray}
for $\nu_{\rm t}<\nu<\nu_{\rm max}$, where $\nu_{\rm t}$ is the turnover frequency
below which the radiation is beamed into a solid angle significantly larger than
$\Omega$. Figure~\ref{fig2} shows this flux for the Crab (PSR~B0531$+$21)
and the two powerful pulsars in the LMC: J0537$-$6910 (which may also 
show $\gamma$-ray flares \cite{saitoetal17}) and B0540$-$69.

The blue dots in this figure indicate the value of $\nu F_\nu$ at
$\nu_{\rm max}$, assuming that $f=1$ and that the pulsar wind remains
undisturbed when propagating in the direction of the observer out to
the point at which $\sigma=1$, at radius $a_{\rm L}r_{\rm L}/2$. 
This is clearly unrealistic for most pulsars; for the Crab $a_{\rm L}r_{\rm L}\approx 4\,$pc. 
Assuming, instead, that the wind, and, therefore, the
inductive acceleration process, terminates roughly at the radius
suggested by x-ray observations \cite{kargaltsevpavlov08}, the maximum
value migrates to the position of the orange points. In this case, a
filling factor $f\approx0.1$ gives an adequate fit to the spectrum of
the powerful flare observed in April 2011\cite{buehleretal12}. The
position of the $\nu_{\rm t}$ in this figure is arbitrarily set to
$80\,$MeV, and the spectrum at lower photon energy is calculated
assuming the electron beam diverges in the nebula to fill a cone of
opening angle given by (\ref{divergence}).  In addition,
Fig.~\ref{fig2} shows the value of $\nu F_\nu$ assuming $r_{\rm
  t.s.}=a_{\rm L}r_{\rm L}/2$ for PSR~J0205$+$6449 (in 3~C~58) and
PSR~B1957$+$20 (the Black Widow). In the former case, the termination
shock projected onto the plane of the sky appears to lie much closer
to the pulsar \cite{kargaltsevpavlov08}, but in the latter its
position is not well constrained.

\section{Summary}
Inductive acceleration provides an attractive scenario
that can be applied to flares in various astrophysical objects.
It does not attempt to explain 
why the supply of charged particles to small parts of a
magnetically dominated, relativistic outflow should suffer
interruptions, but, instead, opens up the possibility of 
studying the physics of
the electromagnetic cascades that are responsible for producing these
charges.  Such cascades are thought to be
non-stationary, both in pulsars and in black hole magnetospheres
\cite{ceruttibeloborodov17,levinsonsegev17}, but their 
spatial and temporal structure remains unclear. In addition, the theory makes
firm predictions on the frequency dependence of the variation timescale,
which should be $\propto 1/\nu^2$, and on 
the polarization of the emission, which, in the pulsar case, should be that 
of synchrotron radiation in the predominantly toroidal nebular field.

\acknowledgments 
This research was supported by a grant from the GIF,
the German-Israeli Foundation for Scientific Research and Development.

\providecommand{\href}[2]{#2}\begingroup\raggedright\endgroup

\end{document}